\documentclass[preprint,trackchanges]{aastex63}

\received{}
\revised{}
\accepted{}
\submitjournal{ApJL}

\graphicspath{{./}{figures/}}

\usepackage{amsmath, bm}

\newcommand{\Ab}{{\bm A}}

\newcommand{\Bb}{{\bm B}}

\newcommand{\Ap}{{\bm A}_{\rm p}}
\newcommand{\Bp}{{\bm B}_{\rm p}}

\newcommand{\nb}{{\bm n}}

\newcommand{\rb}{\hat{\bm r}}

\newcommand{\dlb}{\,\mathrm{d}{\bm l}}

\newcommand{\dS}{\,\mathrm{d}S}
\newcommand{\dV}{\,\mathrm{d}V}
\newcommand{\nablah}{\nabla_{\rm h}}

\begin{document}

\title{The Minimal Helicity of Solar Coronal Magnetic Fields}

\correspondingauthor{Anthony R. Yeates}
\email{anthony.yeates@durham.ac.uk}

\author[0000-0002-2728-4053]{Anthony R. Yeates}
\affiliation{Department of Mathematical Sciences \\
Durham University, Durham, DH1 3LE, UK}

\begin{abstract} % MAX 250 WORDS
Potential field extrapolations are widely used as minimum-energy models for the Sun's coronal magnetic field. As the reference to which other magnetic fields are compared, they have -- by any reasonable definition -- no global (signed) magnetic helicity. Here we investigate the internal topological structure that is not captured by the global helicity integral, by splitting it into individual field line helicities. These are computed using potential field extrapolations from magnetogram observations over Solar Cycle 24, as well as for a simple illustrative model of a single bipolar region in a dipolar background. We find that localised patches of field line helicity arise primarily from linking between strong active regions and their overlying field, so that the total unsigned helicity correlates with the product of photospheric and open fluxes. Within each active region, positive and negative helicity may be unbalanced, but the signed helicity is only around a tenth of the unsigned helicity. Interestingly, in Cycle 24, there is a notable peak in unsigned helicity caused by a single large active region.
On average, the total unsigned helicity at the resolution considered is approximately twice the typical signed helicity of a single real active region, according to non-potential models in the literature.
\end{abstract}


% MAIN TEXT MAX 3500 WORDS (NOT INCLUDING APPENDICES)
% MAX 5 FIGURES+TABLES
% MAX 50 REFERENCES

%=========================================================
\section{Introduction} \label{sec:intro}
%=========================================================

The potential field source surface (PFSS) model, established in the 1960s \citep{schatten1969,altschuler1969}, remains the baseline against which more sophisticated coronal magnetic field models are compared, holding its own as a first approximation for the heliospheric magnetic structure even in the era of \textit{Parker Solar Probe} \citep{badman2020}. Satisfying both $\nabla\cdot\Bp=0$ and $\nabla\times\Bp=\boldsymbol{0}$, the PFSS field $\Bp$ minimizes magnetic energy in the region $r_0<r<r_1$ among all magnetic fields that match a given distribution of $B_r$ on $r=r_0$ and satisfy $B_\theta=B_\phi=0$ on $r=r_1$ \citep{priest}. Typically $r_0$ is the solar surface and $r_1$ is fixed somewhere between $1.5r_0$ and $3r_0$. By minimizing magnetic energy, the PFSS model describes the simplest coronal magnetic structure consistent with (radial component) magnetogram observations at $r=r_0$.

Over the solar cycle, the PFSS corona -- and thus the minimal possible complexity of the real corona -- varies significantly in structure, in turn driving variations in the heliospheric magnetic topology \citep{wang2003,wang2014}. A natural question to ask, therefore, is how to quantify the complexity of this minimal magnetic structure at any given time. One approach is to count topological features such as magnetic null points \citep{cook2009,freed2015,edwards2015b} or separators \citep{platten2014} that divide the corona into regions of differing magnetic connectivity \citep{longcope2005}. Where they extend to the $r_1$ boundary, these connectivity regions determine the origins of different regions of the solar wind, and \citet{scott2019} have recently developed an automated technique for such a partitioning of the so-called S-web \citep{antiochos2011}.

The other approach -- pursued here -- is to describe the PFSS topology in terms of magnetic flux linkage, or helicity integrals. For any magnetic field $\Bb=\nabla\times\Ab$, the magnetic helicity
\begin{equation}
    h(V_t)=\int_{V_t}\Ab\cdot\Bb\dV
    \label{eqn:h}
\end{equation}
is well-known to be an ideal-magnetohydrodynamic invariant in any co-moving subvolume $V_t$ bounded by magnetic surfaces with $\Bb\cdot\nb=0$ (where $\nb$ is the surface normal). It measures the net linking between magnetic flux within $V_t$, and thus characterises the topology of $V_t$ \citep[\textit{e.g.},][]{pevtsov2014,moffatt}. Because the $r_0$ and $r_1$ boundaries are not magnetic surfaces, the coronal volume $V$ ($r_0<r<r_1$) cannot be divided into magnetically-closed subvolumes. But we can divide it in such a way that the only non-magnetic subvolume boundaries lie on $r_0$ and/or $r_1$. Then the individual helicity $h(V_t)$ of any of the subvolumes $V_t$ can change only by evolution of $B_r$ on $r_0$ and/or $r_1$ \citep{demonulin2009}, not by ideal motions inside $V$, even if the latter deform the internal boundaries between the subvolumes.

As it stands, the definition \eqref{eqn:h} does not immediately make sense for magnetically open (sub)volumes because it then depends on the choice of $\Ab$. In Section \ref{sec:def} we will make a specific choice $\Ap$ for our potential field that (i) is as small as possible in that it minimises the integral of $|\Ab|^2$, and (ii) ensures that $h(V)=0$ over the whole volume. The latter is a natural requirement for the minimum-energy field, and accords with the relative helicity that is often used for non-potential fields \citep{berger1984,finn1985,moraitis2018}.

The interesting thing, and the premise of this Letter, is that fixing $\Ab=\Ap$ globally in this way does not imply $\Ap\cdot\Bp=0$ throughout $V$. As we will show, even potential fields may contain subvolumes with $h(V_t)\neq 0$. The lack of electric currents means that these subhelicities cannot arise from local twisting of magnetic field lines, so they must arise from mutual linking between different active regions and/or the overlying large-scale field. We will see the importance of the latter in Sections \ref{sec:bmr} and \ref{sec:cycle}. The presence of distinct connectivity regions is familiar from the aforementioned topological studies, and their mutual linkage -- and therefore subhelicity -- is effectively forced by the distribution of $B_r$ on $r=r_0$ \citep[cf.][]{bourdin2018}.  We comment on the possible significance of this minimal helicity content in Section \ref{sec:conclusion}.

%=========================================================
\section{Definitions} \label{sec:def}
%=========================================================

\subsection{Vector potential}

Any potential field $\Bp$ in $V$ with no net flux through $r=r_0$ (or consequently through $r=r_1$) may be written as  $\Bp=\nabla\times\Ap$ where $\Ap$ is the unique vector potential determined by the conditions $\nabla\cdot\Ap=0$ and $\Ap\cdot\rb=0$ throughout $V$ \cite[\textit{e.g.},][]{berger1988}. One way to find this vector potential is to write
\begin{equation}
    \Ap(r,\theta,\phi) = \nabla\times\big[P(r,\theta,\phi)\rb\big]
    \label{eqn:ap}
\end{equation}
and find the potential $P(r,\theta,\phi)$ by solving the two-dimensional Poisson equation $\nablah^2 P = -B_r$ on each surface of constant $r$. Another is to solve the Poisson equation only on $r=r_0$, then integrate radially \citep[the so-called DeVore-Coulomb gauge;][]{amari2013,yeates2016,moraitis2018} to find
\begin{equation}
    \Ap(r,\theta,\phi) = \frac{r_0}{r}\Ap(r_0,\theta,\phi) + \frac{1}{r}\int_{r_0}^r\Bp(r',\theta,\phi)\times\rb\,r'\,\mathrm{d}r'.
    \label{eqn:devore}
\end{equation}

This particular choice of vector potential is the ``simplest possible'' in that it minimises $\int_V|\Ab|^2\dV$ among all possible vector potentials \citep[cf.][]{gubarev2001}, as well as being a minimiser of $\int_{\partial V}|\Ab\times\rb|^2\dS$ on the boundary \citep[as advocated by][]{yeates2018}. As such, it is appropriate for defining the minimal field line helicity content of a potential field. 

\subsection{Helicity measures} \label{sec:h}

We will use the finest possible subdivision of $V$: infinitesimal magnetic flux tubes surrounding every magnetic field line. Denoting such a tube of radius $\epsilon$ around a field line $L$ by $V_\epsilon(L)$, and the tube's magnetic flux by $\Phi_0(V_\epsilon(L))$, we consider the field line helicity 
\begin{equation}
    \mathcal{A}(L) = \lim_{\epsilon\to 0}\frac{\int_{V_\epsilon(L)}\Ap\cdot\Bp\dV}{\Phi_0(V_\epsilon(L))},
    \label{eqn:lim}
\end{equation}
where the normalisation is needed to give a finite limit. The properties and meaning of field line helicity were first discussed in detail by \citet[][see also \citealp{aly2018}]{berger1988}, who noted \eqref{eqn:lim} as an alternative to the simpler formula
\begin{equation}
    \mathcal{A}(L) = \int_L\Ap\cdot\dlb,
    \label{eqn:flh}
\end{equation}
which is more convenient for calculations \citep{yeates2016,moraitis2019}. 

In a potential field -- which can contain no closed or ergodic field lines -- it follows from \eqref{eqn:lim} that $h(V)$ may be written as a flux-weighted integral of $\mathcal{A}$ over $\partial V$, with
\begin{equation}
    \frac{1}{2}\int_{\partial V}\mathcal{A}|B_r|\dS = \int_V\Ap\cdot\Bp\dV = 0.
    \label{eqn:intabr}
\end{equation}
The factor half arises because each field line hits the boundary twice (here $\partial V$ includes both the inner and outer boundaries), and the second integral vanishes for our choice of $\Ap$. In this sense, $\mathcal{A}$ decomposes the total helicity into a distribution of invariants for each field line. Each one is topologically meaningful because it can change only by motion of the field line endpoints on $\partial V$, or by reconnection within $V$ \citep{berger1988}. In fact, in a highly-conducting plasma, reconnection tends to redistribute $\mathcal{A}$ between field lines, rather than destroy it \citep{russell2015}.

As we will see, the individual field line helicities $\mathcal{A}(L)$ do not vanish in general, even though $\Bp$ is a potential field.
As an overall measure of the field line helicity content of a given potential field, we will use the ``unsigned helicity''
\begin{equation}
    \overline{H} = \frac{1}{2}\int_{\partial V}|\mathcal{A}B_r|\dS.
    \label{eqn:absh}
\end{equation}

%=========================================================
\section{Single bipolar magnetic region} \label{sec:bmr}
%=========================================================

\begin{figure*}
    \centering
    \includegraphics[width=\textwidth]{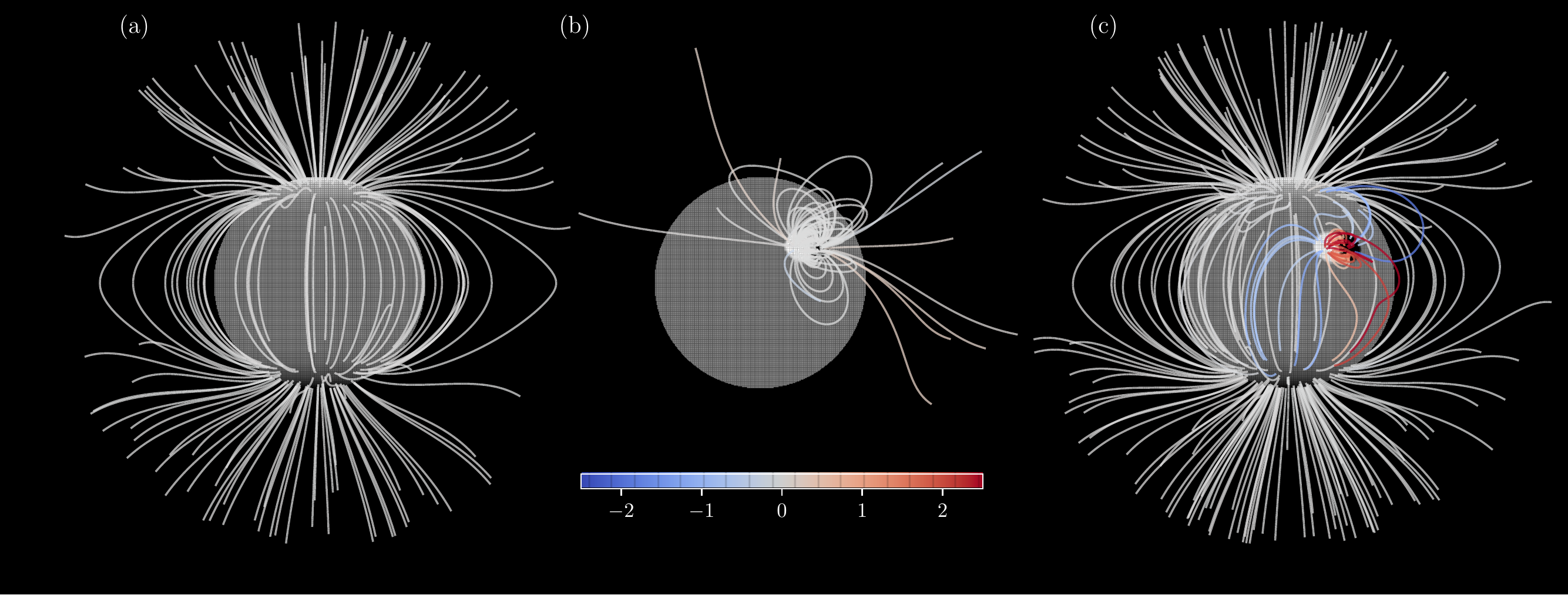}
    \caption{PFSS extrapolations for (a) a dipolar field $B_r(r_0) \sim\cos^7\theta$; (b) a localised BMR (defined in Appendix \ref{app:bmr}); and (c) a superposition of the two (with $0^\circ$ BMR tilt). The peak field strength of the BMR is $50\,\mathrm{G}$ compared to $5\,\mathrm{G}$ for the dipolar field. Selected magnetic field lines are colored blue/red by $\mathcal{A}|B_r|$ (in units of $10^{22}\,\mathrm{Mx}^2\,\mathrm{cm}^{-2}$ and using the larger of the two endpoint $|B_r|$ values). Panel (c) is the top left (tilt $0^\circ$) case in Figure \ref{fig:bmr2d}. }
    \label{fig:bmr3d}
\end{figure*}

Before studying data-driven potential-field extrapolations, it is instructive to consider the field line helicity of a single bipolar magnetic region (BMR).  Figure \ref{fig:bmr3d} shows three PFSS extrapolations, for (a) a dipolar background field, (b) the single BMR, and (c) their superposition. All extrapolations in this paper are computed on a regular grid of $60\times 180\times 360$ points in $(\log(r/r_0), \cos\theta,\phi)$ coordinates, using the author's finite-difference code \citep{yeates2018code}. To calculate $\mathcal{A}$, we first determine $\Ap$ from $\Bp$ using a finite-difference version of \eqref{eqn:devore} and a fast-Poisson solver for $P(r_0,\theta,\phi)$. A second-order Runge-Kutta method is then used to integrate $\Ap$ along magnetic field lines.

On its own, the dipolar background field (Figure \ref{fig:bmr3d}a) has $\mathcal{A}= 0$ on all field lines. To see this, note that axisymmetry of $\Bp$ implies that $P$ is also axisymmetric, so that $\Ap$ has only a $\phi$-component by \eqref{eqn:ap}. But since $B_{{\rm p}\phi}=0$ we have $\Ap\cdot\Bp=0$ throughout $V$.
With no background field, the BMR (Figure \ref{fig:bmr3d}b) also has vanishing $\mathcal{A}$ on every closed field line, but this is due to symmetry: opposite values of $\Ap\cdot\Bp$ are encountered as such a field line undergoes equal displacement toward the BMR and away from it. The open field lines at either end of the BMR have a net displacement toward (or away from) the BMR, so have non-zero (but small) $\mathcal{A}$. The combined field in Figure \ref{fig:bmr3d}(c), however, has much more field line helicity.

\begin{figure*}
    \centering
    \includegraphics[width=\textwidth]{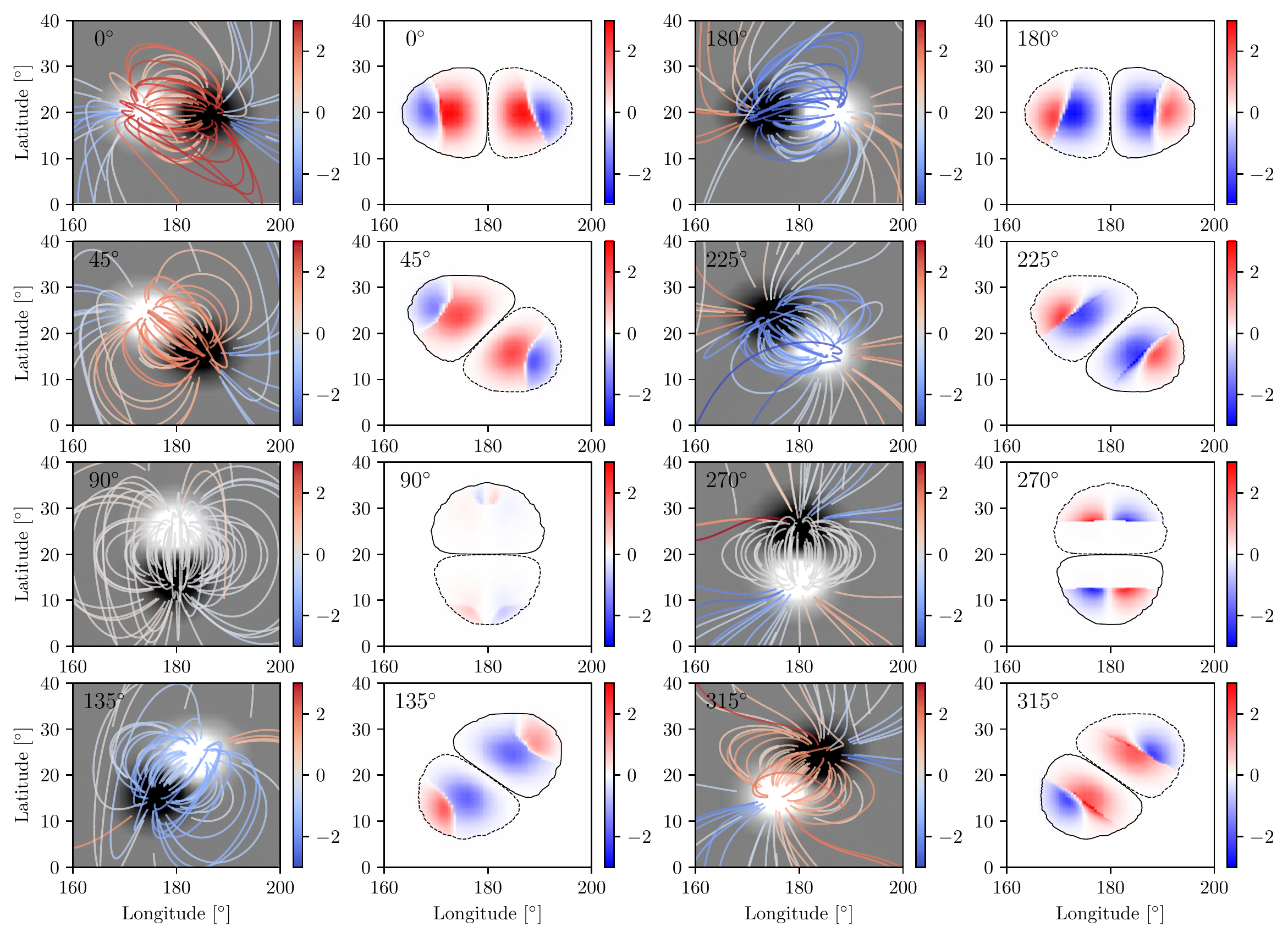}
    \caption{Field line helicity for different orientations of a BMR, labelled by tilt angle in degrees. In each case, the left panel shows $B_r$ on $r=r_0$ (white positive, black negative) with magnetic field lines colored by $\mathcal{A}|B_r|$ (using the larger of the two endpoint $|B_r|$ values). The right panel shows the corresponding distribution of $\mathcal{A}|B_r|$ on $r=r_0$, with $B_r=\pm 1\,\mathrm{G}$ contours in black. Units of $\mathcal{A}|B_r|$ are $10^{22}\,\mathrm{Mx}^2\,\mathrm{cm}^{-2}$. The dipolar background field is always positive to the north and negative to the south.}
    \label{fig:bmr2d}
\end{figure*}

Aside from an overall scaling with magnetic field strength, the distribution of $\mathcal{A}$ in the combined field depends primarily on the orientation (tilt angle) of the BMR. This is shown by Figure \ref{fig:bmr2d}, where the distribution of $\mathcal{A}|B_r|$ is plotted for 8 different orientations of a BMR at the same location (central latitude $20^\circ$ North). The ``helicity content'' -- as measured by $\overline{H}$ -- is maximised at either $0^\circ$ tilt (top left) or $180^\circ$ tilt (top right), when the majority of the BMR flux is perpendicular to the overlying dipolar field. It is minimised at $90^\circ$ when the BMR is aligned with the dipolar field, and is only a little larger at $270^\circ$ when it is anti-aligned.
The sign of $\mathcal{A}$ in each part of the BMR depends on the direction of the East-West magnetic field component relative to the overlying field. In the tilt $0^\circ$ BMR, for example, the closed field lines connecting the two BMR polarities have positive $\mathcal{A}$, whereas the field lines at the extremities connecting elsewhere have negative $\mathcal{A}$. Changing the polarity of the BMR (tilt $180^\circ$) reverses this pattern. For tilt $90^\circ$ or $270^\circ$ symmetry means that the closed BMR field lines have no net East-West displacement, so no net $\mathcal{A}$. 

In fact, the tilt $0^\circ$ BMR shown in Figure \ref{fig:bmr2d} has a net positive helicity. Defining the BMR by the $\pm 1\,\mathrm{G}$ contour on $r=r_0$, we estimate the signed integral of $\mathcal{A}|B_r|$ over the closed field lines connecting within this region to be $1.7\times 10^{42}\,\mathrm{Mx}^2$, while that over field lines originating in this region and connecting elsewhere (either open or closed) is $-1.4\times 10^{42}\,\mathrm{Mx}^2$. The net helicity from the BMR region (inside the $1\,\mathrm{G}$ contour) is therefore $+0.3\times 10^{42}\,\mathrm{Mx}^2$. This is balanced by a net negative contribution from the closed dipolar field lines, but note that it is an order of magnitude smaller than the  unsigned helicity $\overline{H}\approx3.5\times 10^{42}\,\mathrm{Mx}^2$. For the combined field, integrated over the BMR and the dipole, we find  $\overline{H}\approx4.7\times 10^{42}\,\mathrm{Mx}^2$. (All of these estimates used field lines traced from a grid on $r=r_0$ at twice the resolution of the $\Ap$ and $\Bp$ fields. Computing $\mathcal{A}|B_r|$ by summing over these  same field lines, we obtain  $\int_V\Ap\cdot\Bp\dV=-0.07\times 10^{42}\,\mathrm{Mx}^2$, suggesting that $\overline{H}$ is accurate to about $1\%$.) In summary, significant unsigned helicity requires significant BMR flux perpendicular to the overlying field, and the signed helicity of the BMR is approximately $10\%$ of its unsigned helicity.

%=========================================================
\section{Solar cycle evolution} \label{sec:cycle}
%=========================================================

The same numerical code has been used to compute PFSS extrapolations and $\mathcal{A}$ distributions using synoptic line-of-sight magnetogram data from the Helioseismic and Magnetic Imager \citep[HMI,][]{schou2012} on \textit{Solar Dynamics Observatory}. We use the radial component, pole-filled maps in the \texttt{hmi.synoptic\_mr\_polfil\_720s} series \citep{sun2018}, for Carrington Rotations CR2098 (2010 June) to CR2226 (2020 February).
The maps were prepared by (i) applying a smoothing filter of the form $\mathrm{e}^{-b_0l(l+1)}$ to the spherical harmonic coefficients; (ii) mapping to the computational grid using cubic interpolation; and (iii) correcting flux balance. The grid resolution was fixed at $60\times180\times360$ but we tried increasing the smoothing from $b_0=2\times 10^{-5}$ to $1\times10^{-4}$ and $5\times 10^{-4}$. The flux balance was corrected by multiplicative scaling of both the positive and negative regions to their original mean. Before correction, the maps had varying levels of signed flux up to about 5\% of their unsigned flux. However, the signed flux in any given map before correction showed no correlation with the approximately 1\% signed helicity ($H$) found after correction, consistent with the latter arising solely from numerical error in the subsequent calculation. The source surface was fixed at $r_1=2.5\,R_\odot$.

\begin{figure*}
    \centering
    \includegraphics[width=\textwidth]{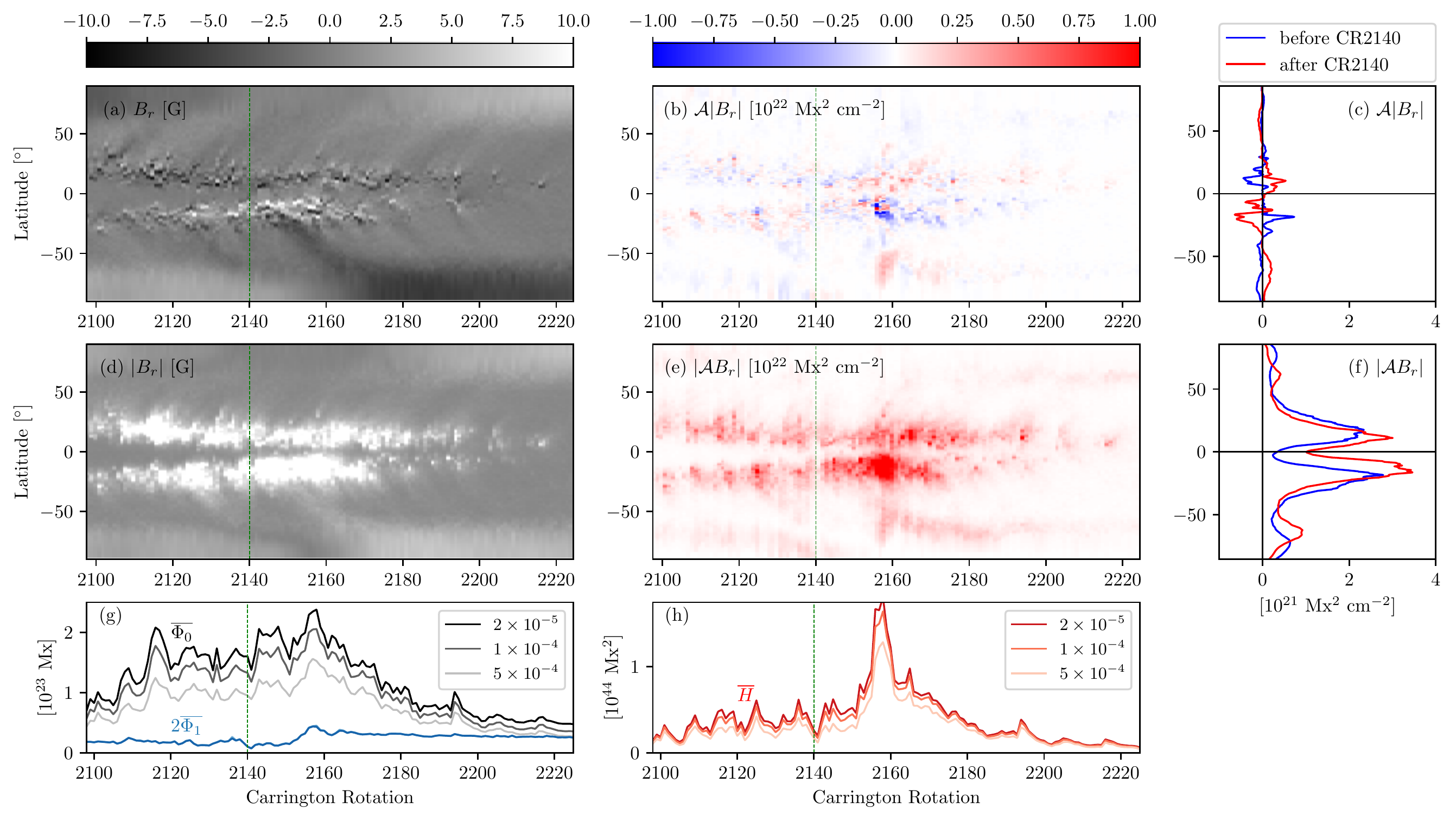}
    \caption{Results of the Cycle 24 PFSS extrapolations. Panels (a) and (b) show longitudinal averages of $B_r$ and $\mathcal{A}|B_r|$ on $r=r_0$ as functions of time and latitude. Panels (d) and (e) show similar averages for $|B_r|$ and $|\mathcal{A}B_r|$. Panels (c) and (f) show time averages of (b) and (e), respectively, separately for the periods before and after the dipole reversal time at CR2140 (dashed green line in the other plots). Finally, as functions of time, panel (g) shows the total unsigned fluxes $\overline{\Phi_0}$ and $\overline{\Phi_1}$ and panel (h) shows $\overline{H}$ (for the different levels of magnetogram smoothing -- the other panels show only the $b_0=1\times10^{-4}$ results).}
    \label{fig:bfly}
\end{figure*}

The left column of Figure \ref{fig:bfly} shows how the magnetic flux evolves over latitude and time in this PFSS model. Panel (g) shows the total (unsigned) fluxes through the inner and outer boundaries, defined as
\begin{equation}
    \overline{\Phi_0} = \frac12\int_{r=r_0}|B_r|\dS, \qquad \overline{\Phi_1} = \frac12\int_{r=r_1}|B_r|\dS.
\end{equation}
Notice in Figure \ref{fig:bfly} that $\overline{\Phi_0}$ depends on the smoothing $b_0$ but the open flux $\overline{\Phi_1}$ does not, as it is controlled by only the lowest few spherical harmonic degrees \citep{wang2014}. This particular solar cycle does not show a sharp peak in $\overline{\Phi_0}$ but there is quite a sharp peak in $\overline{\Phi_1}$ around CR2156-8.

The other columns of Figure \ref{fig:bfly} then summarize the resulting field line helicity. The stackplots in (b) and (e) show longitude averages of $\mathcal{A}|B_r|$ and $|\mathcal{A}B_r|$, respectively, and the time averages of these same quantities are shown in (c) and (f). From (f), we see that the unsigned helicity is predominantly located in the active region belts, and is about ten times the size of the signed helicity in (c), as we found for the single BMR in Section \ref{sec:bmr}. In the active region belts -- between latitudes $\pm30^\circ$ -- the time-averaged signed helicity in (c) has opposite sign in each hemisphere, and also has opposite sign before and after reversal of the Sun's polar field (blue versus red curves). This suggests that the dominant contribution is from the linking of active regions with the overlying dipolar field.

\begin{figure*}
    \centering
    \includegraphics[width=\textwidth]{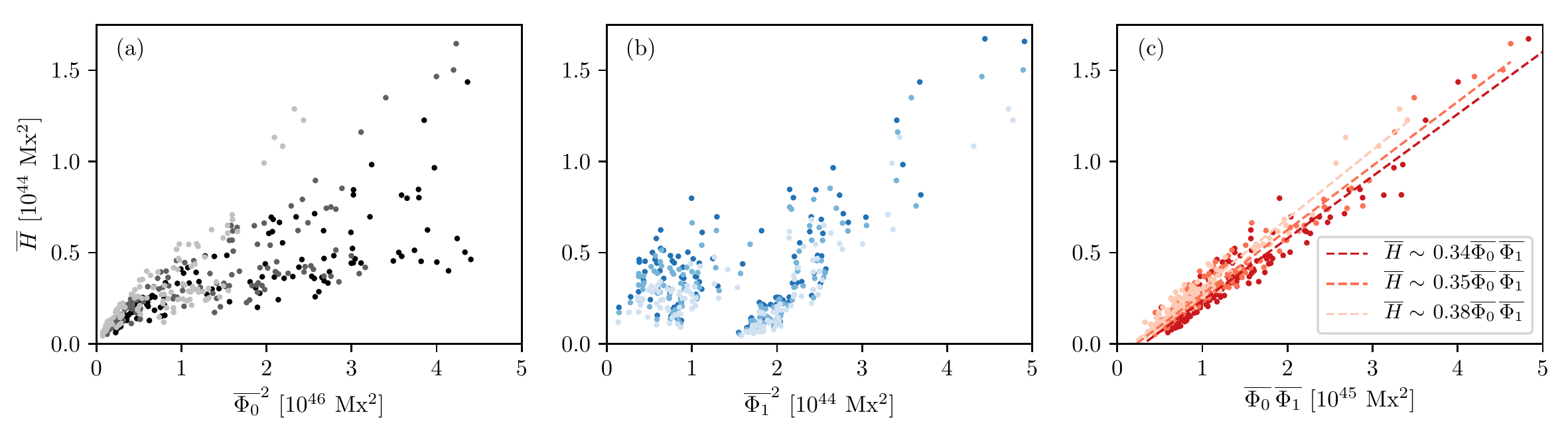}
    \caption{Scatter plots of $\overline{H}$ against (a) $\overline{\Phi_0}^2$,  (b) $\overline{\Phi_1}^2$, and (c) $\overline{\Phi_0}\,\overline{\Phi_1}$ for all Cycle 24 extrapolations. Models with different smoothing $b_0$ are shown by different weights (as per Figure \ref{fig:bfly}). There is no direct relationship in (a) or (b), but least-squares linear fits are shown in (c).}
    \label{fig:cor}
\end{figure*}

Figure \ref{fig:bfly}(h) shows the overall $\overline{H}$ as a function of time -- obtained by integrating (e) over latitude. We find that it correlates most strongly not with $\overline{\Phi_0}^2$ or $\overline{\Phi_1}^2$, but with their product $\overline{\Phi_0}\,\overline{\Phi_1}$ (Figure \ref{fig:cor}). This likely arises because $\overline{\Phi_1}$ itself correlates with the amount of overlying dipolar field above active regions, since both are determined by low-order spherical harmonics. So Figure \ref{fig:cor} further supports the idea that the dominant contribution to field line helicity is linking between active regions and the overlying field.

\begin{figure}
    \centering
    \includegraphics[width=0.5\textwidth]{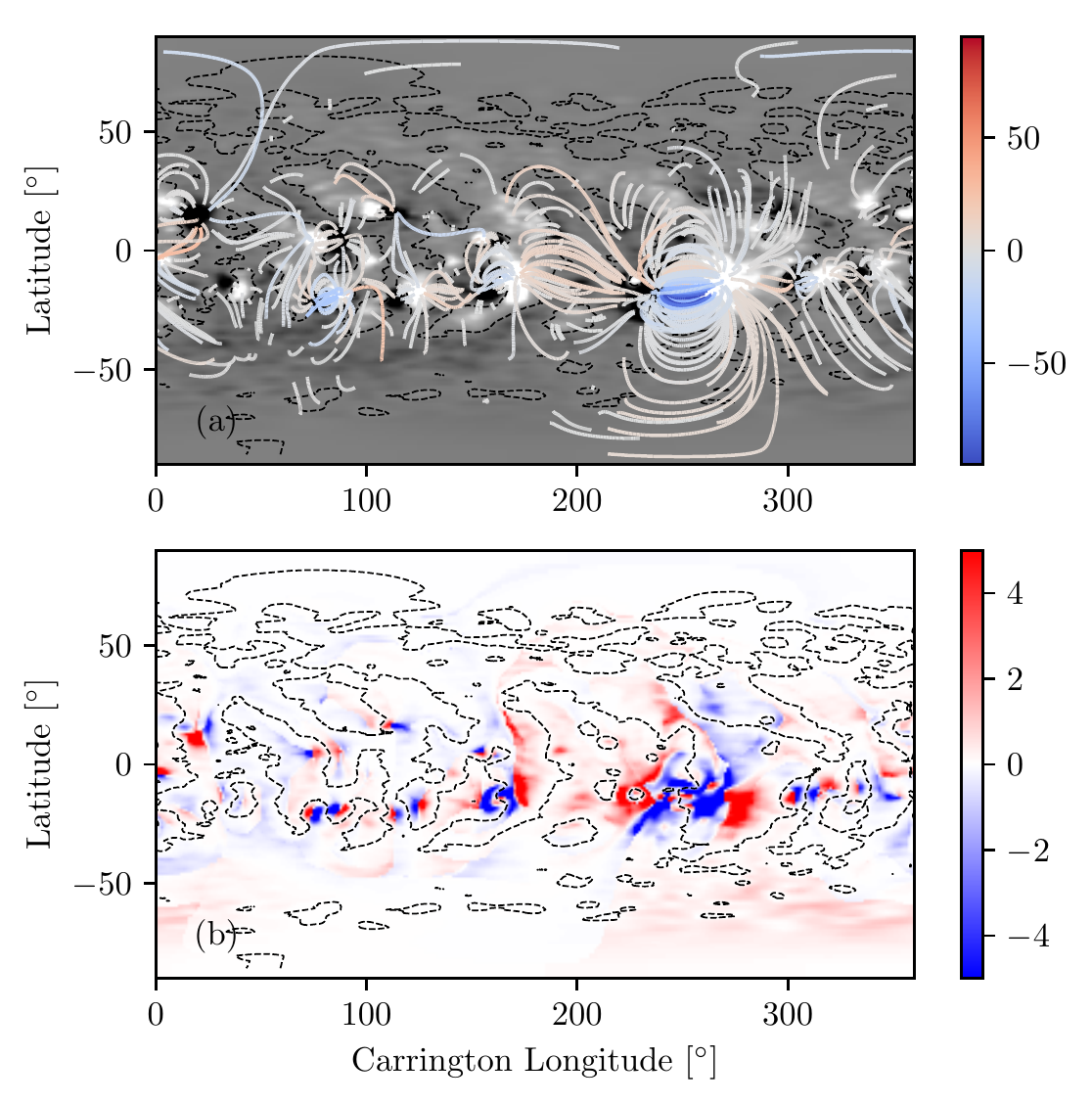}
    \caption{PFSS model for CR2157 ($b_0=1\times10^{-4}$), showing the large active region at longitude $250^\circ$. Panel (a) shows $B_r$ on $r=r_0$ in grayscale (white positive, black negative), along with magnetic field lines colored by $\mathcal{A}|B_r|$ (using the larger of the two endpoint $|B_r|$ values). Panel (b) shows $\mathcal{A}|B_r|$ on $r=r_0$ in blue/red. In both cases the blue/red color scale is in units of $10^{22}\,\mathrm{Mx}^2\,\mathrm{cm}^{-2}$; for clarity it is capped at $\pm 5\times 10^{22}$ in (b) but covers the full range in (a). The dashed black line in both panels shows the zero contour of $B_r$ on $r=r_0$. %An animated version of this figure is available. The video begins at CR2098 and ends at CR2226, with realtime duration of 11 seconds. It places this active region into context as one of the larger active regions within the full solar cycle.
    }
    \label{fig:cr2157}
\end{figure}

The substantial peak in $\overline{H}$ around CR2157-8 is interesting: it arises from a single Southern-hemisphere active region -- the ``Great Solar Active Region'' NOAA 12192 \citep{sun2015}, which was the largest since 1990 \citep{nagy2017}. Shown in Figure \ref{fig:cr2157}, this region has net negative helicity in our PFSS model because it emerges after the polar field reversal with positive leading polarity (following Hale's law).  Its pattern corresponds roughly to the $135^\circ$ case in Figure \ref{fig:bmr2d}.
Although the real region likely had significant free energy not captured in our PFSS model \citep{sun2015}, it did indeed emerge with negative helicity. This is suggested both by the chirality of extreme-ultraviolet loops in the centre of the region and by estimates of current-helicity in the region \citep{mcmaken2017}. It is also suggested by Figure 3 of \citet{pipin2019}, who use HMI vector magnetograms without extrapolation to map the local helicity density $\Ab\cdot\Bb$ on $r=r_0$. (Incidentally, those authors chose the same gauge for $\Ab\times\rb$ on $r=r_0$ as for our $\Ap$, but their $\Ab$ has an additional radial component due to the non-potentiality of the real magnetic field.) 
We reiterate that this peak in $\overline{H}$ in CR2157 does not simply arise because of the peak in $\overline{\Phi}_0$. There is a similar peak in $\overline{\Phi}_0$ in CR2116 of slightly lower magnitude ($1.8\times 10^{23}\,\mathrm{Mx}$ instead of $2.0\times 10^{23}\,\mathrm{Mx}$ for $b_0=1\times 10^{-4}$), but no significant peak in $\overline{H}$ at that time, mainly because $\overline{\Phi}_1$ was weaker than in CR2157 ($0.97\times 10^{22}\,\mathrm{Mx}$ instead of $2.2\times 10^{22}\,\mathrm{Mx}$).

%=========================================================
\section{Discussion} \label{sec:conclusion}
%=========================================================

How do our minimal helicities obtained in Section \ref{sec:cycle} compare to the real corona, which contains non-potential magnetic energy above the minimal PFSS level? It is not yet possible to give a definitive answer, since the non-potential structure of the global coronal field -- particularly outside of active regions -- remains poorly constrained \citep{yeates2018b}. However, we can compare to rough estimates from the literature. The time-averaged  $\overline{H}$ over the whole dataset in Section \ref{sec:cycle} (with $b_0=2\times10^{-5}$) is $4.0\times 10^{43}\,\mathrm{Mx}^2$.  
This is roughly twice the average (relative) helicity content of a significant active region, which observations and data-driven models suggest to be around $2\times 10^{43}\,\mathrm{Mx}^2$ \citep{devore2000,bleybel2002,bobra2008,pevtsov2008,georgoulis2009}. 

Notice that $2\times 10^{43}\,\mathrm{Mx}^2$ is an order of magnitude larger than the signed helicity of our potential BMR in Section \ref{sec:bmr}. This non-potential helicity can arise from emergence of non-potential magnetic structures, from post-emergence footpoint motions within the active region, or from large-scale shearing of the region by (primarily) differential rotation. The latter acts even on an idealised BMR \citep{devore2000}. As shown by \citet{hawkes2019}, the spatial pattern of helicity injection from differential rotation in a BMR is rather different from the patterns of PFSS field line helicity seen in Figure \ref{fig:bmr2d}, showing instead a characteristic north-south pattern of positive and negative injection \citep[see also][]{pipin2020}.

Estimates also exist for the helicity in interplanetary magnetic clouds, which have originated from the corona. Our mean $\overline{H}$ is roughly ten times the helicity of a typical interplanetary magnetic cloud \citep{demoulin2016}, although the Halloween 2003 event was estimated to remove as much as $2\times 10^{44}\,\mathrm{Mx}^2$ from the Sun \citep{lynch2005}. In a magneto-frictional model, \citet{lowder2017} found that erupting flux ropes removed, on average, $2.6\times 10^{43}\,\mathrm{Mx}^2$, a more substantial fraction of $\overline{H}$ in the model.

Putting these estimates together suggests that the unsigned helicity of the PFSS model is not entirely insignificant. Of course, being the minimum energy field, this helicity cannot be released in eruptions unless the pattern of $B_r$ on $r=r_0$ simplifies. This did happen, of course, during Cycle 24 after the peak seen in Section \ref{sec:cycle}. We finish by remarking that, even if the unsigned helicity of the PFSS model is modest, its basic field line helicity pattern may well be imprinted in the real non-potential field. For example, numerical simulations show that this minimal helicity can act as a seed for amplification by photospheric shearing motions \citep{yeates2016}, ultimately explaining the pattern of positive and negative helicity observed in highly sheared filament channels \citep{yeates2009}. Even helicity arising from the magnetic structure on small scales will tend to collect in these filament channels \citep{knizhnik2017}, ultimately leading to flares or eruptions.

\acknowledgments

This work was supported by The Leverhulme Trust (grant PRG-2017-169) and the UK STFC (grant ST/S000321/1), and benefited from the discussions of the ISSI International Team on Magnetic Helicity in Astrophysical Plasmas. The author thanks P. Wyper and two anonymous reviewers for significantly improving the paper. The \textit{SDO} data are courtesy of NASA and the \textit{SDO}/HMI science team.


\vspace{5mm}
\facilities{SDO (HMI)}


%\software{astropy \citep{2013A&A...558A..33A},  
%          Cloudy \citep{2013RMxAA..49..137F}, 
%          SExtractor \citep{1996A&AS..117..393B}
%          }



\appendix

\section{Functional form for the BMR} \label{app:bmr}

We obtain the expression for $B_r(r_0,\theta,\phi)$ of our bipolar magnetic region (BMR) by first defining an untilted BMR located on the equator at longitude $\phi=0$ and latitude $\lambda=\pi/2-\theta=0$. This has the form 
\begin{equation}
B_r(r_0,\lambda,\phi) = -B_0\frac{\phi}{\rho}\exp\left[-\frac{\phi^2 + 2\lambda^2}{\rho^2}\right],
\end{equation}
where in this paper we take the heliographic separation angle $\rho=5^\circ$ and choose $B_0$ to give the peak field strength $50\,\mathrm{G}$. The magnetic flux of each polarity is then $5.15\times10^{21}\,\mathrm{Mx}$, consistent with a moderately-sized active region. To create a BMR centred at $(\lambda_0,\phi_0)$ with tilt angle $\gamma_0$, we apply a sequence of three rotations: (i) through angle $-\gamma_0$ around the $x$-axis; (ii) through angle $-\lambda_0$ around the $y$-axis, and (iii) through angle $+\phi_0$ around the $z$-axis, where $x$, $y$, $z$ are Cartesian coordinates defined by $x = \cos\phi\cos\lambda$, $y = \sin\phi\cos\lambda$, $z=\sin\lambda$.


%\bibliography{yeates}{}
%\bibliographystyle{aasjournal}



\end{document}